\documentclass{baltzer}
\begin{document}
\begin{frontmatter}
%
\title{Resonant Scattering of Muonic Hydrogen Atoms and Dynamics
of Muonic Molecular Complex
}       

\author[a,b]{M. C. Fujiwara\thanks{e-mail:Makoto.Fujiwara@cern.ch}}
\author[c]{A.~Adamczak},
\author[d]{J.M.~Bailey},
\author[e]{G.A.~Beer},
\author[f]{J.L.~Beveridge},
\author[g]{M.P.~Faifman},
\author[h]{T.M.~Huber},
\author[i]{P.~Kammel},
\author[j]{S.K.~Kim},
\author[k]{P.E.~Knowles},
\author[l]{A.R.~Kunselman},
\author[m]{V.E.~Markushin},
\author[f]{G.M.~Marshal},
\author[e]{G.R.~Mason},
\author[k]{F.~Mulhauser},
\author[f]{A.~Olin},
\author[m]{C.~Petitjean},
\author[n]{T.A.~Porcelli},
\author[o]{J.~Wo\'zniak},
\author[p]{J.~Zmeskal}
\address[a]{University of British Columbia, Canada}
\address[b]{University of Tokyo, Japan}
\address[c]{Inst.~Nucl.~Physics Cracow, Poland}
\address[d]{Chester Technology, UK}
\address[e]{University of Victoria, Canada}
\address[f]{TRIUMF, Canada}
\address[g]{Kurchatov Institute, Russia}
\address[h]{Gustavus Adolphus College, USA}
\address[i]{University of Illinois at Urbana-Champaign, USA}
\address[j]{Jeonbuk National University, S.~Korea}
\address[k]{Universit\'{e} de Fribourg, Switzerland}
\address[l]{University of Wyoming, USA}
\address[m]{PSI, Switzerland}
\address[n]{University of Northern British Columbia, Canada}
\address[o]{Inst.~Phys.~Nucl.~Tech., Poland}
\address[p]{Austrian Academy of Sciences, Austria}

%
%

 \runningauthor{M.C. Fujiwara et al.}
\runningtitle{Resonant Scattering of Muonic Hydrogen Atoms}
%
\begin{abstract}  
Resonant scattering of muonic hydrogen atoms via back decay of
molecular complex, a key process in the understanding of
epithermal muonic molecular formation, is analyzed. The
limitations of the effective rate approximation are discussed and
the importance of the explicit treatment of the back decay is
stressed. An expression of the energy distribution for the
back-decayed atoms is given.

\end{abstract}
\end{frontmatter}
Direct measurements of epithermal resonant molecular formation,
recently reported by the TRIUMF Muonic Hydrogen
Collaboration~\cite{fujiw00,porce01,marsh01}, required detailed
considerations of processes which were not previously
well-studied. Resonant scattering of muonic atoms via back decay
of the muonic molecular complex (MMC) is one such process. Despite
substantial theoretical efforts in improving the accuracies of
muonic hydrogen scattering cross sections, so far little attention
has been paid to the back decay process as a scattering mechanism
of the muonic atom (except for the spin flip in the $d\mu d$ system).
It is the purpose of this paper to emphasize the importance of
the resonant scattering and the associated MMC dynamics with the
hope of stimulating further theoretical studies.

Resonant formation of MMC, %
$\mu a^F + DX _{\nu _i K_i} \rightarrow  [(d\mu a)^S_{11} x
ee]_{\nu _f K_f}$%
taking place with the rate $\lambda ^{SF}_ {\nu _i K_i, \nu _f
K_f} (E _{\mu a})$, is generally followed by competing processes
of either stabilization with the effective rate $\tilde{\lambda}
_f$ leading to fusion, or back decay to $\mu a^{F} + DX _{\nu _i'
K_i'}$ with the width $\Gamma ^{SF}_{\nu _f K_f, \nu _i' K_i'}$.
Here $a=d,t$, $x=p,d,t$, and $X=H,D,T$. $F$ is the hyperfine state
of $\mu a$, $S$ is the spin of $d\mu a$, and $\nu _i K_i$, $\nu _f
K_f$ are the vibrational and rotational quantum number of $DX$ and
MMC respectively. Our notations follow those of
Ref.~\cite{faifman}, but we explicitly account for the possibility
of (de)excitation of $DX$ upon back decay $\nu _i' K_i' \neq \nu
_i K_i$, {\it i.e.,} resonant (de)excitation, whose importance will
become apparent below.

In the analysis of conventional $\mu$CF experiments, an effective
renormalized formation rate~\cite{faifman} has been widely used,
into which the effect of the back decay is absorbed as:
\begin{equation}\label{eq:eff}
  \tilde{\lambda}^F _{d\mu a} = \sum _{\nu _f, K_f, S}
    W^{SF}_{\nu _f K_f}
  \sum _{K_i} \omega _{K_i} \lambda ^{SF}_ {\nu _i K_i, \nu _f K_f}
    ,
\end{equation}
where $W^{SF}_{\nu _f K_f} = \tilde{\lambda} _f /(\tilde{\lambda}
_f+ \sum _{\nu _i' K_i'} \Gamma ^{SF}_{\nu _f K_f, \nu _i' K_i'})$
is the fusion probability, and
 $\omega _{K_i}$ is the initial $K_i$ population.
 We observe that even in the case where the transport of muonic
atoms can be neglected, at least one of the following criteria
must met in order to justify the effective rate approximation of
Eq.~\ref{eq:eff} in describing fusion yields: (a) trivial
condition that the back decay probability $(1-W^{SF}_{\nu _f K_f})
\ll 1$, (b) rapid (compared to MMC formation) re-thermalization of
$\mu a$ in the equilibrium condition, or (c) negligible change in
$\mu a$ energy in {\it lab} frame before and after the back decay.
For example, the condition (a) is satisfied for $d\mu t$ formation
at low energies, while the condition (b) applies for $d\mu d$ (at
least at high densities), and (c) may be possible in condensed
matter if recoil-less processes dominate.  If none of the above
criteria are satisfied, the $\mu a$ can be removed from the
resonance regions affecting the kinetics, which was indeed the
case in the experiment of Refs.~\cite{fujiw00,porce01,marsh01}.
In simplified model calculations, where one interaction, either
of resonant scattering or potential scattering, is assumed to be
sufficient to remove $\mu a$ from the resonance region, the
effective rate approximation overestimates the fusion yield by
$(\lambda ^{SF}_{\nu _iK_i,\nu _fK_f} + \lambda _{scat}
)/(W^{SF}_{\nu _f K_f}\lambda ^{SF}_{\nu _iK_i,\nu _fK_f}+\lambda
_{scat})$ where $\lambda _{scat}$ is the potential scattering rate,
compared to the explicit inclusion of resonance scattering
channel~\cite{fujiw99}. For $\mu t + D_2$ at resonance peak
energy, this factor is as large as $\sim 1.5$.

A more realistic estimate of the effect of resonant scattering
requires the accurate $\mu a$ energy distribution after back
decay, which depends on the details of the MMC dynamics including:
(i) MMC recoil from $\mu a$ impact upon its formation, (ii)
thermalization of MMC center of mass motion in collision with the
target medium, (iii) collisional relaxation/excitation of MMC
ro-vibrational states ($\nu _f K_f \rightarrow \nu _f'K_f'$), (iv)
MMC decay with possible $DX$ excitation, and (v) $DX$ recoil upon
MMC decay.

Apart from trival, but often a neglected effect of the kinematics
(i,v), the cross section for [$(d\mu t)dee] + D_2$ elastic collisions
has been calculated by Padial {\it et al.}~\cite{padia88} for
relatively low energies; its extrapolation to epithermal energies
suggests a value of about $3\times 10^{-15}$ cm$^2$. The corresponding
collision rate $\sim 2\times 10^{13}$ s$^{-1}$ is an order of
magnitude larger than the MMC decay rate $10^{12}$ s$^{-1}$, hence
substantial thermalization can be expected at high densities.

Calculations of rotational transitions in the MMC were reported by
Ostrovskii and Ustimov~\cite{ostro80}, and by Padial {\it et
al.}~\cite{padia88b} for the case of thermal equilibrium targets.
Ostrovskii and Ustimov estimate relaxation rates of the order of
10$^{13}$ s$^{-1}$ (a value used in Ref.~\cite{faifman}), while
Padial {\it et al.}, claiming higher accuracy, give $\sim 0.3
\times 10^{13}$ s$^{-1}$ at 300~K (rates are normalized to liquid
hydrogen density).

To date, there are no accurate calculations available for
vibrational quenching of MMC, except for a rough estimate by
Lane~\cite{lane}, who gives $10^7$ s$^{-1}$ at room temperature, a
rate much slower than other processes. He predicts, however,
increasing rates for higher temperature and increasing $\nu$.
Future accurate calculations of this process is highly desirable,
as the consequence of non-negligible quenching would also have a
significant impact on other aspects of molecular formation, such
as the fusion rate $\tilde{\lambda}_f$.

Given the initial condition $\nu _i, K_i, E_{\mu a}$, the
approximate energy distribution of back decayed $\mu a$, $f(E_{\mu
a}')$, can be estimated from the following
expression~\cite{fujiw99}:
\begin{eqnarray}
f^F(E_{\mu a}') &= & \sum _{\nu _f' K_f' S} h^{SF}_{\nu _i K_i}
(\nu _f' K_f'; E_{\mu a}) \sum_{v_i',K_i'} g^{SF}_{\nu _f K_f}
(\nu_{i}',K'_{i})   \nonumber
\\ & \times &  \int dE_u D(E_u)
I\left[ E'_{\mu a} - E'_{\nu _i K_i} (\nu _f K_f S;\nu '_i
K'_i)-E_u \right], \label{eq:Edist}
\end{eqnarray}
 where
\begin{equation}
  E'_{\nu _i K_i}(\nu _f K_f S;\nu '_i K'_i)  =
\frac {M_{DX}}{M_{DX}+M_{\mu a}} \left[\epsilon _{res} (\nu
_{f}K_{f}) - \Delta E_{v_i K_i,v'_i K'_i} \right]
\label{eq:energy}
\end{equation}
is the decay Q-value for the specific channel $(\nu _f, K_f, S)$
$\rightarrow$ $(\nu '_i, K'_i, S)$, with $\Delta E_{v_i K_i,v'_i
K'_i} = - \left[ E (\nu' _{i}K'_{i}) - E (\nu _{i}K_{i}) \right]$
being the binding energy difference between the initial and final
state of $DX$ ({\it i.e.,} (de)excitation energy of $DX$ due to
the resonant scattering). $I[\Delta]$ is the resonance intensity
profile for detuning $\Delta $ ({\it e.g.,} $I[\Delta] = \delta
(\Delta)$ in the classical Vesman model), while $D(E_u)$ is the
Doppler broadening distribution due to the motion of MMC at the
time of back decay ({\it e.g.,} Gaussian distribution with the
width $\sigma = \sqrt{4kTM_{\mu a}/M_{MMC}}$ for thermalized MMC
if $E'_{\mu a} \gg kT$).
\begin{equation}
g^{SF}_{\nu f K_f}(\nu '_i,K'_i) = \frac{\displaystyle \Gamma
^{SF}_{\nu _{f}K_{f}, \nu ^\prime _{i}K^\prime _{i}} }
{\displaystyle \sum_{\nu ^\prime   _{i},K^\prime _{i}} \Gamma
^{SF}_{\nu _{f}K_{f}, \nu ^\prime _{i}K^\prime _{i}} }
\end{equation}
is the branching ratio for the decay into the state $(\nu '_i,
K'_i, S)$, given the MMC state of $(\nu _f, K_f, S)$,
and $h^{SF}_{\nu _i K_i} (\nu _f K_fS; E_{\mu a})$ is the
ro-vibrational population of MMC at the time of back decay, given
the initial condition $\nu _i, K_i$ and $E_{\mu a}$.
A full evaluation of Eq.~\ref{eq:Edist} would require, in addition
to the back decay matrix elements, a solution of kinetics equation
involving all the competing processes; some limiting cases were
considered in Ref.~\cite{fujiw99}.
We note that in Refs. \cite{somov,jeitl}, back-decayed $\mu t$ is
suggested to have a thermal energy distribution of the target
temperature, but we find this not be the case, even if MMC
translational motion and/or ro-vibrational states are completely
thermalized (see Eq.\ref{eq:energy}).

In our analysis for Ref.~\cite{fujiw00}, $E_{\mu t}'$ was varied
between 1 meV to 0.3 eV in Monte Carlo calculations~\cite{huber99}
explicitly taking into account the resonant scattering (an
improved version of earlier calculation \cite{marku96}) to
phenomelogically investigate its effect, and some 7\% difference
in fusion yield was observed, giving a non-negligible contribution
to the total systematic uncertainties. Although this error is not
overwhelming, a substantial improvement in the accuracy of
resonant formation measurements would require, among others,
detailed understanding of resonant scattering processes and MMC
dynamics, the first step of which has been illustrated in this
report.


%
\end{document}